\documentstyle[aps,prd]{revtex}
\begin{document}
\draft

\title{\bf On Duality in the Born-Infeld Theory}
\author{A. Khoudeir and Y. Parra}
\address{Centro de Astrof\'{\i}sica Te\'orica, Departamento de
F\'{\i}sica,
Facultad de Ciencias, Universidad de los Andes, M\'erida, 5101,
Venezuela.}
\maketitle
\begin{abstract}
The $SL(2,R)$ duality symmetric action for the 
Born-Infeld theory in terms of two potentials, coupled with non-trivial 
backgroud fields in four dimensions is established.
This construction is carried out in detail by analysing the hamiltonian 
structure of the Born-Infeld theory.
The equivalence with the usual Born-Infeld theory is shown.
\end{abstract}


Nowadays the concept of duality is widely recognized by its 
unifying role in physics. The five known different superstring theories 
are now unified by duality in the framework of M Theory 
(see for example \cite{sch1}). The simplest case where duality appears
are 
the Maxwell's equations without sources, interchanging the equations of
motions 
and the Bianchi identities. Schwarz and Sen \cite{ss} have developed a
method 
to raise duality symmetry at the level of the action but at the price 
of losing the explicit Lorentz invariance. The classical and quantum 
equivalence with electromagnetism has been well established \cite{alv}, 
\cite{gir}, \cite{ggrs}. Earlier, Deser and Teitelboin \cite{dt}  
noticed that the Maxwell theory in its hamiltonian formulation is 
invariant under non-local dualty transformations. Moreover, several 
attempts have been made at conciliating duality symmetry 
with Lorentz invariance \cite{kptsb}. On the other hand, the Born-Infeld 
theory \cite{bi}, initially conceived as an alternative for
electromagnetism, 
has recently received considerable attention because the world volume
action 
of a D-brane is described by a kind of non-linear Born-Infeld action
\cite{polas}. Several aspects of the duality symmetry in the Born-Infeld 
theory have been developed recently \cite{123}, \cite{beng}.
In particular, Perry and Schwarz \cite{sch2} proposed a non-manifestly 
Lorentz invariant Born-Infeld action for a self-interacting self-dual 
antisymmetric tensor field in $D=6$. Afterward, Pasti, Sorokin and Tonin 
presented a manifestly covariant formulation of this action \cite{pst2}, 
from which Berman \cite{berman}, after dimensional reduction to four
dimensions 
and breaking the Lorentz symmetry, obtained an action for the
Born-Infeld 
theory coupled with the axion and dilaton fields with a $Z(2)$ symmetry.

In this article, we will study the Born-Infeld theory both pure and
coupled 
with the axion and dilaton fields from the point of view of their
hamiltonian 
structures in four dimensions, setting up the formulation $\it {a \ la}$ 
Schwarz-Sen of the $SL(2,R)$ duality invariant Born-Infeld action.
Our results extend those obtained by Berman when the axion field is 
involved, making evident the $SL(2,R)$ duality invariance. 
We will show that our results lead to the Born-Infeld theory after 
the elimination of one of two potentials.

We will start with the Born-Infeld theory without non-trivial background
fields.
Despite its highly non-linear character, the abelian gauge 
Born-Infeld theory whose action is given by
\begin{equation}
I = \int d^4 x \left [1 - \sqrt{-det(\eta_{mn} + F_{mn})}\right] ,
\end{equation}
describes the casual and ghost free propagation of massless spin-1
fotons 
like Maxwell's theory. In order to perform the hamiltonian analysis it
will 
be convenient to introduce an auxiliary field $v$ \cite{tse} and rewrite 
down the Born-Infeld action as 
\begin{equation}
I = \int d^4 x \left [1 - \frac{1}{2}v(1 + \frac{1}{2}F^{mn}F_{mn} -
\frac{1}{16}
( {\cal{F}} ^{mn} F_{mn})^2) - \frac{1}{2v}\right].
\end{equation}
We have used the fact that in four dimensions 
$\mid det(\eta_{mn} + F_{mn})\mid $ = $(1 + \frac{1}{2}F^{mn}F_{mn} -
\frac{1}{16}
({\cal{F}} ^{mn} F_{mn})^2)$ 
and ${\cal{F}} ^{mn} \equiv \frac{1}{2}\epsilon^{mnpq}F_{pq}$ is the 
dual of the field strength $F_{mn} = \partial_m A_n - \partial_n A_m$.
The canonical momenta are found to be
\begin{equation}
\pi _i = \frac{\delta {\cal L}}{\delta \dot A_i} = vD_{ij}E_j, \quad 
\pi^0 = \frac{\delta {\cal L}}{\delta \dot A_0} = 0 , \quad 
p = \frac{\delta {\cal L}}{\delta \dot v} = 0,
\end{equation}
where $E_i \equiv F_{0i}$, $B_i \equiv  \epsilon^{ijk}\partial_j A_k $
and 
$D_{ij} \equiv \delta_{ij} + B_i B_j$. Our flat metric signature is 
$(-1,+1,+1,+1)$. $\phi ^1 \equiv \pi^0 \approx 0$ and $\phi ^2 \equiv p 
\approx 0$ are the primary constraints. Since $D$ is a non-singular
matrix 
($detD = 1 + \vec{B}.\vec{B} \neq 0$) there exist the inverse of $D$ and 
from eq.(3) we obtain that 
\begin{equation}
\dot A_i = \frac{1}{vdetD}\left [\delta_{ij}detD - B_i B_j \right]\pi_j
+ 
\partial_i A_0 .
\end{equation}
 
It is straightforward to compute the canonical hamiltonian
\begin{equation}
H_o = \frac{1}{2v}(\vec{\pi}.\vec{\pi}+ 1) + \frac{v}{2}(\vec{B}.\vec{B}
+ 1) 
-\frac{1}{2vdetD}(\vec{\pi}.\vec{B})^2 - 1 - A_0\partial_i \pi_i .
\end{equation}

Requiring consistency of the primary constraint $\phi ^1$, the Gauss's 
constraint $\phi ^3 \equiv \partial_i \pi_i \approx 0$ follows
inmediately, 
while the consistency of the constraint $\phi ^2 $ lead us to determine
the 
value of the auxiliary field $v$:
\begin{equation}
v = \frac{1}{detD}\sqrt{(\vec{\pi}.\vec{\pi}+ 1)(\vec{B}.\vec{B}+ 1) - 
(\vec{\pi}.\vec{B})^2}.
\end{equation}

It is worth recalling that the canonical hamiltonian is invariant under 
non-local duality transformation \cite{dt}
\begin{equation}
\delta A_i = \theta \nabla ^{-2}\epsilon^{ijk}\partial_j \pi_k, \quad 
\delta \pi _i = \theta\epsilon^{ijk}\partial_j A_k ,
\end{equation}
whenever the auxiliary field transforms as 
\begin{equation}
\delta v = \frac{2\theta}{detD}v(\vec{\pi}.\vec{B}).
\end{equation}

Eliminating the auxiliary field $v$ using eq.(6), the well known 
non-linear canonical hamiltonian \cite{beng}, \cite{dirac} is recovered
\begin{equation}
H_o = \sqrt{1 + \vec{\pi}.\vec{\pi} + \vec{B}.\vec{B} + 
(\vec{\pi} {\bf x} \vec{B}).(\vec{\pi} {\bf x} \vec{B})} 
- 1 - A_0\partial_i \pi_i .
\end{equation}

The set of constraints are first class ones and we can write the BRST 
invariant generating functional (choosing the Coulomb gauge) for the 
Born-Infeld theory as
\begin{equation}
Z = \int DA_i D\pi_i Dc D\bar{c} 
\delta (\partial_i \pi_i) \delta (\partial_i A_i) exp (iI_{eff}),
\end{equation}
where $c$ and $\bar{c}$ are the pair of ghost-antighost fields
associated 
to the first-class constraint $\phi^3$ and the effective action is 
\begin{equation}
I_{eff} = \int d^4 x [\pi_i \dot{A_{i}} - H_o - i\bar{c}\nabla^{2}c],
\end{equation}
which is invariant under duality transformations (7). $c$ and 
$\bar{c}$ are inert under duality and since they are decoupled from 
the pair $\pi_i$ and $A_i$, we will omit their presence from now on.
Moreover, we will assume that the Gauss's constraint is solved, then
both 
$\vec{\pi}$ and $\vec{A}$ are transverse.

Following the ideas developed in references \cite{dght}, 
\cite{ggrs} we introduce two new fields (renaming $A_i^{1} \equiv A_i$ 
and $\pi_i^{1} \equiv \pi_i$)
\begin{equation}
A_i^{2} \equiv \epsilon^{ijk}\nabla ^{-2}\partial_j \pi_k^{1}, \quad 
\pi_i^2 \equiv \epsilon^{ijk}\partial_j A_k^{1},
\end{equation}
with the goal to achieve the duality transformations
(${\cal{L}}_{12} = 1 = -{\cal{L}}_{21}$):
\begin{equation}
\delta A_{i}^{\alpha} = {\cal{L}}_{\alpha \beta}A_{i}^{\beta}, \quad
\delta \pi_{i}^{\alpha} = {\cal{L}}_{\alpha \beta}\pi_{i}^{\beta},
\end{equation}
generated by an 
abelian Chern-Simons term: $G = \int d^3 x \epsilon^{ijk}, 
A_{i}^{\alpha}\partial{j}A_{k}^{\alpha}$.
The action (11) is rewriten down as 
\begin{equation}
I = \int d^4 x \left [\frac{1}{2}\pi_{i}^{\alpha}\dot{A_{i}}^{\alpha} 
- \sqrt{1 + B_{i}^{\alpha}B_{i}^{\alpha} + \frac{1}{4}(\epsilon^{ijk}
B_{j}^{\alpha}{\cal{L}}_{\alpha \beta}B_{k}^{\beta})^2} + 1 \right],
\end{equation}
where $\vec{B}^1 = \vec{\nabla} {\bf x} \vec{A}^1$ and 
$\vec{B}^2 = \vec{\nabla} {\bf x} \vec{A}^2 = \vec{\pi}^1$. 
This action is just the canonical form for 
the following action, in terms of two potentials $\it{a \ la}$ 
Schwarz-Sen, for the Born-Infeld theory
\begin{equation}
I = - \int d^4 x \left [\frac{1}{2}B_{i}^{\alpha}{\cal{L}}_{\alpha
\beta}E_{i}^
{\beta} + \sqrt{1 + B_{i}^{\alpha}B_{i}^{\alpha} +
\frac{1}{4}(\epsilon^{ijk}
B_{j}^{\alpha}{\cal{L}}_{\alpha \beta}B_{k}^{\beta})^2} - 1 \right]  ,
\end{equation}
clearly invariant under duality transformations.
This result coincides with the previously obtained by Berman
\cite{berman} 
in the absence of non-trivial background fields. In the weak 
approximation, where we neglected the higher terms in
$(B_{i}^{\alpha})^2 $, 
the Schwarz-Sen action for Maxwell theory is obtained.

Now, we will consider the Born-Infeld theory coupled with an axion and 
a dilaton, described by the following action:
\begin{equation}
I = \int d^4 x \left [1 - \sqrt{-det(\eta_{mn} + 
e^{-\frac{1}{2}\phi}F_{mn})} + 
\frac{1}{4}\psi {\cal{F}} ^{mn} F_{mn}\right].
\end{equation}
It is well known that the equations of motion derived from this action 
are $SL(2,R)$ duality invariant \cite{123} (we have omitted the kinetic
terms 
for the dilaton and the axion which are $SL(2,R)$ duality invariant). 
Besides this action is $SL(2,R)$ self-dual 
\cite{tse}. Introducing the auxiliary field $v$, we rewrite down the
action as 
\begin{equation}
I = \int d^4 x \left [1 - \frac{1}{2}v(1 + \frac{1}{2}e^{-\phi}
F^{mn}F_{mn} - 
\frac{1}{16}e^{-2\phi}( {\cal{F}} ^{mn} F_{mn})^2) - \frac{1}{2v} + 
\frac{1}{4}\psi {\cal{F}} ^{mn} F_{mn} \right].
\end{equation}
In that case, the canonical momentum associated to $A_{i}$ is
\begin{equation}
\pi _i = \frac{\delta {\cal L}}{\delta \dot A_i} = ve^{-\phi}D_{ij}E_j + 
\psi B_i ,
\end{equation}
where now $D_{ij} \equiv \delta _{ij} + e^{-\phi}B_{i}B_{j}$, its 
inverse given by $D_{ij}^{-1} \equiv detD ^{-1} [\delta _{ij}detD 
- e^{-\phi}B_{i}B_{j}]$ and $detD = 1 + e^{-\phi}\vec{B}.\vec{B}$.
The canonical hamiltonian is found to be 
\begin{eqnarray}
H_o &=& \frac{1}{2v}(e^{\phi}\vec{\pi}.\vec{\pi}+ 1) + 
\frac{v}{2}(e^{-\phi}\vec{B}.\vec{B} + 1) -
\frac{1}{2vdetD}(\vec{\pi}.\vec{B})^2\\ \nonumber 
&-& \frac{1}{vdetD}\psi e^{\phi}(\vec{\pi}.\vec{B})
+ \frac{1}{2vdetD}\psi^2 e^{\phi}\vec{B}.\vec{B} - 1 - A_0\partial_i
\pi_i .
\end{eqnarray}
Preserving the primary constraint $p \approx 0$ the value of the
auxiliary 
field $v$ is determined:
\begin{equation}
v = \frac{1}{detD}\sqrt{(e^{\phi}\vec{\pi}.\vec{\pi}+
1)(e^{-\phi}\vec{B}.\vec{B}+ 1) - 
[(\vec{\pi}.\vec{B})^2 + 2\psi e^{\phi}(\vec{\pi}.\vec{B}) - 
\psi^2 e^{\phi}\vec{B}.\vec{B}]}.
\end{equation}
Putting this back into $H_{o}$, we obtain the non-linear canonical
hamiltonian 
in four dimensions for the Born-Infeld coupled with the dilaton and the
axion 
fields. 
\begin{eqnarray}
\nonumber
H_o &=& \sqrt{1 + e^{\phi}\vec{\pi}.\vec{\pi} + e^{-\phi}\vec{B}.\vec{B}
+ 
(\vec{\pi} {\bf x} \vec{B}).(\vec{\pi} {\bf x} \vec{B}) 
- 2\psi e^{\phi}(\vec{\pi}.\vec{B}) + \psi^2 e^{\phi}\vec{B}.\vec{B}} \\ 
&-& 1 - A_0\partial_i \pi_i .
\end{eqnarray}
Introducing two new field variables like eq.(12), 
after solving the Gauss's constraint, we arrive to the 
following hamiltonian in terms of two potentials
\begin{equation}
H_o = \sqrt{1 + B_{i}^{\alpha}({\cal{L}}^T M {\cal{L}}) _{\alpha
\beta}B_{i}^{\beta} + 
\frac{1}{4}(\epsilon^{ijk}
B_{j}^{\alpha}{\cal{L}}_{\alpha \beta}B_{k}^{\beta})^2} - 1 ,
\end{equation}
where we have introduced the symmetric $SL(2,R)$ matrix
\begin{equation}
M = \left( 
\begin{array}{cc}
 e^{\phi} & \psi e^{\phi}\\ 
 \psi e^{\phi} & e^{-\phi} + \psi^2 e^{\phi} 
\end{array}
\right) ,
\end{equation}
satisfying
\begin{equation}
M = M^T, \quad M{\cal{L}}M^T = {\cal{L}}.
\end{equation}

This is the hamiltonian for the following non-manifestly Lorentz
invariant
Born-Infeld action 
coupled to the axion and dilaton fields, in terms of two potentials
\begin{equation}
I = - \int d^4 x \left [\frac{1}{2}B_{i}^{\alpha}{\cal{L}}_{\alpha
\beta}E_{i}^
{\beta} + \sqrt{1 + B_{i}^{\alpha}({\cal{L}}^T M {\cal{L}})_{\alpha
\beta}B_{i}^{\beta} 
+ \frac{1}{4}(\epsilon^{ijk}
B_{j}^{\alpha}{\cal{L}}_{\alpha \beta}B_{k}^{\beta})^2} - 1\right].
\end{equation}

We propose this action as the generalization of the Born-Infeld 
theory in terms of two potentials, 
invariant under the $SL(2,R)$ duality transformations
\begin{equation}
M \rightarrow \omega^{T} M \omega, \quad 
A_i ^{\alpha} \rightarrow \omega^{T}_{\alpha \beta}A_i ^{\beta} ,
\end{equation}
with
${\cal \omega} = \left( 
\begin{array}{cc}
 a & b\\ 
 c & d 
\end{array}
\right)$ 
$\in SL(2,R)$ ($det\omega = +1$).
Now, we are going to show that after eliminating $A_{i}^{2}$ from the 
action (25), we will obtain the Born-Infeld theory coupled with the
dilaton 
and axion fields in the axial gauge $A_{0}^{\alpha} = 0$. Introducing an 
auxiliary variable $v$, the action (25) is rewritten down as
\begin{eqnarray}
I &=& - \int d^4 x [\frac{1}{2}B_{i}^{\alpha}{\cal{L}}_{\alpha
\beta}E_{i}^
{\beta} + 
\frac{1}{2}v(1 + e^{-\phi}B_{i}^1 B_{i}^1 + e^{\phi}B_{i}^2B_{i}^2 + 
(B_{i}^1 B_{i}^1)(B_{i}^2 B_{i}^2) - (B_{i}^1 B_{i}^2)^2 
- 2\psi e^{\phi}B_{i}^1 B_{i}^2 + \psi^2 e^{\phi}B_{i}^1 B_{i}^1 )\\
\nonumber 
&+& \frac{1}{2v} - 1].
\end{eqnarray}
Independent variations in $A_{i}^2$, lead us to the following equations
of motion
\begin{equation}
\epsilon^{ijk}\partial_{j}[ve^{\phi}(\delta_{kl}(1 + e^{-\phi}B_{i}^1
B_{i}^1 ) 
- e^{-\phi}B_{k}^1 B_{l}^1 )B_{l}^2 - E_{k}^1 - ve^{\phi}\psi B_{k}^1 ]
= 0.
\end{equation}
After using the freedom of gauging the gauge symmetry: $\delta \vec{A}^1
= 
\vec{\nabla} \xi$, we can solve eq. (28) for $B_{i}^2$
\begin{equation}
B_{i}^2 = (detD)^{-1}D_{ij}[v^{-1}e^{-\phi}E_{j}^1 + \psi B_{j}^1]. 
\end{equation}
Substituting this value of $B_{i}^2$ into the action (27), we obtain 
\begin{equation}
I = - \int d^4 x [\frac{1}{2}ve^{-\phi}B_{i}^1 B_{i}^1 -
\frac{1}{2vdetD}e^{-\phi}
E_{i}^1 E_{i}^1 - \frac{1}{2vdetD}e^{-2\phi}(E_{i}^1 B_{i}^1)^2 +
\frac{1}{2}v 
+ \frac{1}{2v} - \psi E_{i}^1 B_{i}^1 - 1].
\end{equation}
Finally, we can eliminate the auxiliary field $v$ through its equation
of motion: 
\begin{equation}
v = (detD)^{-1}\sqrt {1 + e^{-\phi}(B_{i}^1 B_{i}^1 - E_{i}^1 E_{i}^1) -
e^{-2\phi}
(E_{i}^1 B_{i}^1)^2 }
\end{equation}
and putting this back into eq. (30), the Born-Infeld action (eq. (16))
is recovered.

It would be interesting to analyze the quantum aspects of this theory as
well as  
find out its supersymmetric extension. Moreover, it would be
enlightening to
try to construct the covariant version of this formulation. 

\underline{Acknowledgements}:
We would like to thank N. Pantoja for useful discussion and D. Morales 
for reading the manuscript.
One of the authors (AK) would like to thank D. Berman for useful 
suggestions after the preliminary version was completed.

\end{document}